\documentstyle[12pt,psfig]{article}

\newcommand{\apl}{\:^{<}_{\sim}\:}

\def\refe{\par\noindent\hangindent=1.5cm}
\begin{document}
\begin{center}

{\bf \large STABILITY OF THE PROTOGALACTIC CLOUDS: I. FIELD LENGTH IN THE ADIABATIC MODELS}

\vspace{0.5cm}

{\large Milan M. \'Cirkovi\'c} \\
\vspace{0.2cm}
{\it
Astronomical Observatory, Volgina 7, 11000 Belgrade, SERBIA \\
Dept. of Physics \& Astronomy, SUNY at Stony Brook, \\
Stony Brook, NY 11794-3800, USA}
\end{center}

\begin{abstract}
All gasdynamical models for the evolution of gaseous content of 
galaxies assume that cooling from the hot, virialized phase to the 
cold phase occured through some sort of thermal instability. 
Subsequent formation of colder clouds embedded in the hot, rarefied
medium is a well-known process appearing in many astrophysical 
circumstances and environments. The characteristics of the condensed 
clouds depend on the relevant timescales for cloud
formation and disruption due to either collisions or one of the 
operating instabilities.
In this paper, the importance of the Kelvin-Helmholtz instability is investigated for
the clouds forming in huge gaseous haloes of $L_\ast$ galaxies. Recent treatment of
this problem by Kamaya (1997) is extended and a more realistic cooling function employed. 
Results show that the Kelvin-Helmholtz instability proceeds effectively on the same timescale whether we account for self-gravity or not. This has multiple significance, since these objects may have been seen as high-column density absorption line systems against the background QSOs, and probably represent the progenitors of the present-day globular clusters.
\end{abstract}

\section{Introduction}
This is the first in a series of papers intended to discuss stability of metal-poor clouds in pressure equilibrium with the quasi-hydrostatic hot gaseous corona of (proto)galaxies. 
It seems that a sort of consensus exists that a state of virialized gaseous haloes located in the potential wells created mainly by the dark matter preceeded
the formation of galactic subsystems as we know them today (Gott \& Thuan 1976; 
White \& Rees 1978; Rees 1978; Miyahata \& Ikeuchi 1995; Mo \& Miralda-Escud\'e 1996). In the transition from this early stage to later steps necessary for the substructure formation, the major role certainly 
belonged  to the process of cooling of gas. Various cooling scenarios were proposed, but it seems clear that the onset of thermal instability in a hot, virialized plasma led to creation of a
two-phase medium, in which cold clouds are in pressure equilibrium with the ambiental,
collisionally ionized medium. Cold clouds may be photoionized by an ionizing background. In the
gravitational potential of the entire halo (i.e.\ baryonic + non-baryonic matter), cold clouds will certainly tend to fall toward the halo center. 

It is important to note that these physical processes are relevant not only for unobservable
early stages of galaxy evolution, but are operational at later epochs either, as in the
Lyman-limit and metal absorption systems and, most probably, at least some of the lower column density Ly$\alpha$ forest absorbers residing in extended haloes of luminous galaxies (e.g.\ 
Chen et al.~1998). Thus, a related motivation for undertaking research in this direction is the possibility to account for some fraction of the observed population of high-column density QSO absorption systems (Mo \& Miralda-Escud\'e 1996).

\section{Field length in galactic-sized haloes}

The thickness of the boundary-layer where the phase transition of plasma is expected 
to be close to the so-called Field length, which is determined from the balance between the thermal conduction and radiative cooling (Field 1965; Aharonson, Regev \& Shaviv 1994).
It can be written as 
\begin{equation}
\label{flen}
\lambda_F = \sqrt{\frac{\kappa T_h}{n_h^2 \Lambda (T_h)}},
\end{equation}
where $\kappa$ is the "classical" coefficient of thermal conduction. The cooling function for the dominant free-free cooling is calculated in detail in Sutherland \& Dopita (1993). 

The temperature dependence of the coefficient of thermal conduction can be, in the regime considered, approximated as (Cowie \& McKee 1977; Kamaya 1997)
\begin{equation}
\label{flen2}
\kappa (T) = \kappa_0 T^{2.5},
\end{equation}
where the constant is given as $\kappa_0 = 5.6 \times 10^{-7}$ erg s$^{-1}$ K$^{-1}$ cm$^{-1}$. 
From Eqs.~(\ref{flen}) and (\ref{flen2}), explicitly taking into account the galactocentric distance of the considered cloud,  we obtain
\begin{equation}
\label{flen3}
\lambda_F = \frac{7.5 \times 10^{-4}}{n_h(r)}  \sqrt{\frac{T_h^{3.5} (r)}{\Lambda[T_h(r)]}}\; {\rm cm}.
\end{equation}
We see that the Field length is estimated from the physical condition ($T_h$, $n_h$) of the {\it hot} plasma component. Its physical meaning is that the cold component will eventually evaporate
(the timescale is discussed in detail in the Cowie \& McKee [1977] paper) if its size
is smaller than $\lambda_F$. On the other hand, the hot component will tend to condense on
the cold cloud surface if its size is larger than $\lambda_F$. Now, our task is clearly formulated: to establish what is the value of $\lambda_F$ at various galactocentric distances 
in a particular gaseous halo model.

The density and temperature profiles in the adiabatic model (Mo 1994; Mo \& Miralda-Escud\'e 1996) are given by
\begin{equation}
\label{density}
\rho_h(r)= \rho_h(r_c) \left( 1 - K \ln \frac{r}{r_c}  \right)^\frac{3}{2},
\end{equation}
and 
\begin{equation}
\label{temp}
T_h(r)= T_h(r_c) \left( 1 - K \ln \frac{r}{r_c}  \right).
\end{equation}
Notation is the following: $K$ is a dimensionless constant equal to
\begin{equation}
\label{const}
K= \frac{2}{5} \frac{\mu V_c^2}{k_{\rm B} T_h(r_c)},
\end{equation}
$k_{\rm B}$ being the Boltzmann constant, $\mu$ average mass per particle, and
$T_h(r_c)$ is temperature at the cooling radius, by assumption equal to the
virial temperature: $T_h(r_c) \equiv \mu V_c^2/2k_{\rm B}$ (see Waxman \&
Miralda-Escud\'e 1995
for an interesting discussion). For a typical $V_c
\simeq 250$ km s$^{-1}$ corresponding to a $L_\ast$ galaxy, and a metallicity
$Z=0.3 \: Z_\odot$ this temperature is  $T_h(r_c)=4.89 \times 10^6$ K, and the
constant $K$ has the value $K= \frac{4}{5}$. The density at the cooling radius is
obtained by requiring the cooling time at $r_c$ to be equal (at least) to
$t_m$:
\begin{equation}
\label{svi}
\rho_h(r_c)=\frac{5 \mu k_{\rm B} T_h(r_c)}{2 \Lambda[T_h(r_c)] t_m},
\end{equation}
where $\Lambda[T_h(r_c)]$ is the cooling rate evaluated at the temperature of
hot phase at the cooling radius. 

Kamaya (1997) has used a simple analytical approximation for the cooling function in the
bremsstrahlung regime (see also Lepp et al.~1985)
\begin{equation}
\label{sim1}
\Lambda (T) \approx 3.0 \times 10^{-27} T^\frac{1}{2} \; {\rm erg} \; {\rm cm}^3 \; {\rm s}^{-1}.
\end{equation}

\begin{figure}
\psfig{file=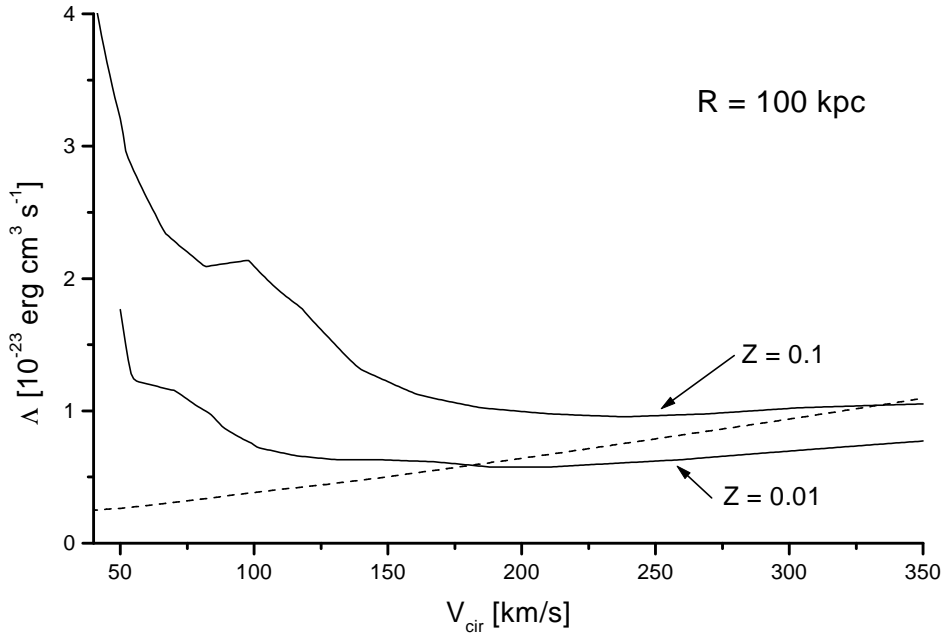,height=10cm}
\caption{Comparison between analytic approximation of Kamaya (1997), represented by dashed
line, and more exact numerical
calculations of Sutherland \& Dopita (1993) for two different metallicities (solid lines) in the
framework of adiabatic halo models. Evaluation is performed at fixed galactocentric distance
of $R=100$ kpc.}
\end{figure}

In Fig.~1, we see the difference between this form of the cooling function and the more
precise numerical values of Sutherland and Dopita (1993). Cooling function for two relevant
metallicities ($0.1 \; Z_\odot$ and $0.01 \; Z_\odot$), in units of 
$10^{-23}$ erg cm$^3$ s$^{-1}$, is represented by solid lines, and
analytic approximation of the eq.~(\ref{sim1}) is shown as the dashed line. The comparison
is made within framework of the adiabatic halo model (although the same applies to any
other physical situation) at the fixed distance $R=100$ kpc from the center of the halo
characterized by the circular velocity $V_{\rm cir}$. The distance of 100 kpc is chosen because
it is far enough from the region of the disk and non-stationary phenomena associated with the
disk-halo connection. Simultaneously, it lies within the virial radius for almost all
halo masses, so that the approximation of quasistationary cooling of gas can be legitimately
applied. 

\begin{figure}
\psfig{file=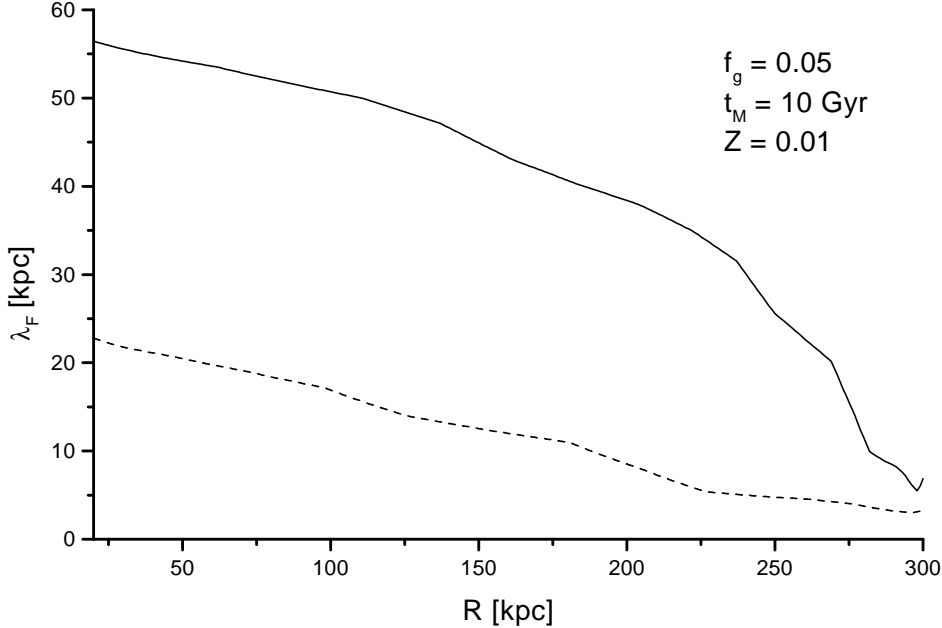,height=10cm}
\caption{Field lengths for clouds at various galactocentric distances in the adiabatic haloes for $V_{\rm cir} =220$ km s$^{-1}$ (solid line) and $V_{\rm cir} =100$ km s$^{-1}$ (dashed line).}
\end{figure}

We see that for circular velocities characteristic for masses of $L_\ast$ galaxies (like the Milky Way), the analytic expression in (\ref{sim1}) is a moderately good approximation
for low-metallicity gas. It remains good within a factor of 2 up to the largest galactic
masses (characterized by $V_{\rm cir} \apl 300$ km s$^{-1}$), but tends to systematically 
underestimate the cooling rate for smaller galactic haloes. This is especially pronounced
in case of dwarf galaxies ($50 \leq V_{\rm cir} \leq 100$ km s$^{-1}$), where the disagreement
is drastic. 

Now we use Eq.~(\ref{flen3}) in conjunction with Eqs.~(\ref{density}) and (\ref{temp}).
The resulting Field lengths are shown in Fig.~2 as functions of galactocentric distances
for the two characteristic values of circular velocity. We see that Field lengths in 
general are quite large; clouds usually envisaged in adiabatic two-phase halo models
(Mo \& Miralda-Escud\'e 1996; Chiba \& Nath 1997) will evaporate through thermal conduction,
on the time scale $t_{\rm evap}$, as discussed by Cowie \& McKee (1977). 

It is immediately clear that large values of the Field length are a consequence of extraordinarily
small density of the hot plasma. It is easy to see that this value must always be low, irrespective of whether we are dealing with huge protogalaxies like those in Kamaya (1997)
picture, or galactic-sized hot haloes of somewhat more modest size. If we perform a thought
experiment and distribute the entire baryonic mass of an $L_\ast$ galaxy, $M_B \sim 5
\times 10^{11}\; M_\odot$ (e.g.~Fields, Freese \& Graff 1998) over the halo of radius $R \sim
100$ kpc, maximal density at the edge of such halo is $\sim 5 \times 10^{-3}$ cm$^{-3}$, and since the adiabatic profile is quite shallow, this can not be very different from our
model values, implying Field lengths of tens of kpc. The existence of collapsed structures, like disk stars or MACHOs, will further increase this lengthscale.

However, if the clouds are contiguous structures of the sizes $\sim 100$ kpc (and still 
surrounded by the hot medium) inferred by Dinshaw et al.~(1994, 1995), they will be stable against evaporation. On the contrary, they will grow through condensations of hot
ambiental plasma.

\section{Application of the Richardson Criterion}

K-H instability which will tend to shred the cold component will be stabilized by self-gravity if the so-called Richardson criterion 
\begin{equation}
\label{rich1}
J > 0.25,
\end{equation}
is satisfied. It is important to emphasize that other means of stabilization (through magnetic fields, for example), are irrelevant of the Richardson criterion, so the breakdown of this criterion does not automatically imply the onset of K-H instability. In the above inequality, the Richardson number $J$ is defined as
\begin{equation}
\label{rich2}
J \equiv - \frac{g}{\rho_c} 
\frac{d\rho / dr}{(dV/dr)^2} \sim 
\frac{g}{\rho_c}  \frac{\rho_c - \rho_h}{V_{\rm cir}^2} \lambda_F.
\end{equation}
The gravity of the cloud at position $r$ is denoted by $g$, and the density of cold clouds by $\rho_c$. Density and velocity gradients, $d\rho / dr$ and $dV/dr$ are evaluated {\it at the boundary layer\/} between the two phases, and corresponding vectors are oriented in such a way as to 
point from the cold to the hot component (e.g. $d \rho /dr <0$). In the second part of this relation, we have used the approximation of Kamaya (1997), in assuming that $d\rho = \rho_h - \rho_c$, $dr = \lambda_F$. For gravitational acceleration, we may write the classical relation
\begin{equation}
\label{rich3}
g = \frac{GM_c}{R_c^2},
\end{equation}
where mass and radius of the cold cloud are denoted by $M_c$ and $R_c$ respectively. Finally, 
we have approximated the relative velocity between the cloud and hot medium by the circular velocity $V_{\rm cir}$. There are several reasons why we should prefer this value to the approximation used by Kamaya (1997) which reduces to the virial velocity of {\it gaseous subsystem} only. The presence of large quantities of dark matter (95\% by mass according to our
fiducial adiabatic model) justifies using its virial velocity as the relevant velocity scale 
within entire halo. Besides, the motions of clouds are likely to be bound from above by terminal
velocity (e.g.~Benjamin \& Danly 1997), which is in any case smaller than the circular velocity
(see also the discussion in Mo \& Miralda-Escud\'e 1996). 

Using all these approximations, we can find Richardson number for two-phase medium of protogalactic or early galactic haloes in the form:
\begin{eqnarray}
\label{fghh}
J & = & \frac{GM_c}{R_c^2} \frac{\lambda_F}{V_{\rm cir}^2} \left( 1 - \frac{\rho_h}{\rho_c} \right)  \approx \nonumber \\
& \approx & 8.9 \times 10^{-4} \! \left( \frac{V_{\rm cir}}{220\; {\rm km\; s}^{-1}} \right)^{-2}\! \!
\left( \frac{M_c}{10^6\; M_\odot} \right)  \! \left( \frac{R_c}{1\; {\rm kpc}} \right)^{-2}  \!
\! \left( \frac{\lambda_F}{10\; {\rm kpc}} \right).
\end{eqnarray}
We have used the fact that adiabatic models and other two-phase pictures predict pressure equilibrium of the two stable thermal phases with $\rho_h / \rho_c \sim 10^{-2}$. Of course, it is not necessary to achieve high precision in any of the terms in order to see that the resulting value is much smaller than the critical value of 0.25.

\section{Discussion}
We have redone the calculations of Kamaya (1997) intended to demonstrate the independence
of the Kelvin-Helmholtz instability of cloud self-gravity. We have improved that work in
two major respects:
\begin{enumerate}
\item
We have worked within the context of a specific, adiabatic halo model of Mo \& Miralda-Escud\'e
(1996), instead of postulating appropriate temperatures and densities.

\item
The more realistic cooling function based on numerical calculations of Sutherland \& Dopita
(1993) was used instead of the analytic approximation.
\end{enumerate}
The main conclusion, however, remains
the same: inclusion of self-gravity in the models of cold clouds formed by thermal instability
does not change their subjection to the Kelvin-Helmholtz instability. This conclusion is valid not only for two-component model of protogalaxies of Miyahata \& Ikeuchi (1995) and Kamaya 
(1996), but also for adiabatic models of galactic haloes of Mo (1994) and Mo \& Miralda-Escud\'e
(1996) intended to explain the Lyman-limit absorption in QSO spectra. 

It is of foremost importance to model physical conditions in the early stages of galactic
history, since that would enable us to understand the origin of the well-known features of
galaxies observable today. To this end, it is necessary to achieve better contact of several, seemingly distinct, fields of astrophysical research. In forthcoming papers of this series, we shall discuss other physical processes affecting cold condensations in galactic-sized haloes and limiting their lifetimes, like the Jeans instability and collision timescales.   

\subsection*{Acknowledgements}

The author is happy to hereby express his gratitude to Dr. Hou Jun Mo for providing a very
useful cooling subroutine. The kind support and encouragement of Milica Topalovi\'c is also
wholeheartedly acknowledged. 

\section*{References}
\addcontentsline{toc}{section}{References}
\refe Aharonson, V., Regev, O. \& Shaviv, N. 1994, ApJ, 426, 621

\refe Benjamin, R. \& Danly, L. 1997, ApJ, 481, 764

\refe Chen, H.-W., Lanzetta, K. M., Webb, J. K. \& Barcons, X. 1998, ApJ, 498, 77 

\refe Chiba, M., \& Nath, B. B. 1997, ApJ, 483, 638

\refe Cowie, L. L. \& McKee, C. F. 1977, ApJ, 211, 135

\refe Dinshaw, N., Impey, C. D., Foltz, C. B., Weymann, R. J. \& Chaffee, F. H. 1994, ApJ, 437, L87

\refe Dinshaw, N., Foltz, C. B., Impey, C. D., Weymann, R. J. \& Morris, S. L. 1995, Nature, 373, 223 

\refe Field, G. B. 1965, ApJ, 142, 531

\refe Fields, B. D., Freese, K. \& Graff, D. S. 1998, NewA, 3, 347

\refe Gott, J. R. III \& Thuan, T. X. 1976, ApJ, 204, 649

\refe Ikeuchi, S. \& Norman, C. A. 1991, ApJ, 375, 479

\refe Kamaya, H. 1996, ApJ, 465, 769

\refe Kamaya, H. 1997, PASJ, 49, 435

\refe Kang, H., Shapiro, P. R., Fall, S. M. \& Rees, M. J. 1990, ApJ, 363, 488

\refe Lepp, S., McCray, R., Shull, J. M., Woods, D. T. \& Kallman, T. 1985, ApJ, 288, 58

\refe Miyahata, K. \& Ikeuchi, S. 1995, PASJ, 47, L37
  
\refe Mo, H. J. 1994, MNRAS, 269, L49  

\refe Mo, H. J. \& Miralda-Escud\'{e}, J. 1996, ApJ, 469, 589 
  
\refe Rees, M. J. 1978, Phys. Scripta 17, 371

\refe Sutherland, R. S. \& Dopita, M. A. 1993, ApJS, 88, 253 

\refe Waxman, E. \& Miralda-Escud\'{e}, J. 1995, ApJ, 451, 451  

\refe White, S. D. M.  \& Rees, M. J. 1978, MNRAS, 183, 341 

\end{document}